\documentclass[lettersize,journal]{IEEEtran}
\usepackage{amsmath,amsfonts}
\usepackage{algorithmic}
\usepackage{algorithm}
\usepackage{array}
\usepackage[caption=false,font=normalsize,labelfont=sf,textfont=sf]{subfig}
\usepackage{textcomp}
\usepackage{stfloats}
\usepackage{url}
\usepackage{verbatim}
\usepackage{graphicx}
\usepackage{cite}
\usepackage{xcolor}
\usepackage[normalem]{ulem}
\usepackage{lineno}

\begin{document}

\title{Development of a Cryogenic Anti-Coincidence System Based on Scintillating Crystals Read Out with Kinetic Inductance Detectors for BULLKID-DM}

\author{%
T.~Lari\textsuperscript{3,4},
G.~Spina\textsuperscript{3,4},
A.~Acevedo-Renter\'{\i}a\textsuperscript{5},
L.~E.~Ardila-Perez\textsuperscript{9},
L.~Bandiera\textsuperscript{6},
C.~Bonomo\textsuperscript{1,2},
M.~Calvo\textsuperscript{7},
M.~Cappelli\textsuperscript{1,2},
R.~Caravita\textsuperscript{10},
F.~Carillo\textsuperscript{4},
F.~Cescato\textsuperscript{6,11},
U.~Chowdhury\textsuperscript{7},
D.~A.~Crovo\textsuperscript{9},
A.~Cruciani\textsuperscript{2},
A.~D'Addabbo\textsuperscript{8},
M.~del~Gallo~Roccagiovine\textsuperscript{1,2},
M.~De~Lucia\textsuperscript{3,4},
G.~Del~Castello\textsuperscript{2},
D.~Delicato\textsuperscript{1,2,7},
F.~Ferraro\textsuperscript{8},
M.~Folcarelli\textsuperscript{1,2},
S.~Fu\textsuperscript{8},
R.~Gartmann\textsuperscript{9},
M.~Grassi\textsuperscript{4},
V.~Guidi\textsuperscript{11,6},
D.~L.~Helis\textsuperscript{12,8},
L.~Malagutti\textsuperscript{6},
A.~Mazzolari\textsuperscript{11,6},
A.~Monfardini\textsuperscript{7},
T.~Muscheid\textsuperscript{9},
D.~Nicol\`o\textsuperscript{3,4},
F.~Paolucci\textsuperscript{3,4},
D.~Pasciuto\textsuperscript{2},
L.~Pesce\textsuperscript{1,2},
C.~Puglia\textsuperscript{4},
D.~Quaranta\textsuperscript{1,2},
C.~M.~A.~Roda\textsuperscript{3,4},
S.~Roddaro\textsuperscript{3,4},
M.~Romagnoni\textsuperscript{6},
G.~Signorelli\textsuperscript{3,4},
F.~Simon\textsuperscript{9},
A.~Tartari\textsuperscript{4},
E.~V\'azquez-J\'auregui\textsuperscript{5},
K.~Zhao\textsuperscript{12,8},
and~M.~Vignati\textsuperscript{1,2}
\\
\IEEEmembership{%
    (1) Dipartimento di Fisica, Sapienza Università di Roma, Roma, Italy, \\
    (2) INFN Sezione di Roma, Roma, Italy, \\
    (3) Dipartimento di Fisica ``Enrico Fermi'', Università di Pisa, Pisa, Italy, \\
    (4) INFN Sezione di Pisa, Pisa, Italy, \\
    (5) Instituto de Física, Universidad Nacional Autónoma de México, Ciudad de México, México, \\
    (6) INFN Sezione di Ferrara, Ferrara, Italy, \\
    (7) Université Grenoble Alpes, CNRS, Grenoble INP, Institut Néel, Grenoble, France, \\
    (8) INFN Laboratori Nazionali del Gran Sasso, Assergi (AQ), Italy, \\
    (9) Institute for Data Processing and Electronics, Karlsruhe Institute of Technology, Eggenstein-Leopoldshafen, Germany, \\
    (10) INFN - TIFPA, Povo (Trento), Italy, \\
    (11) Dipartimento di Fisica e Scienze della Terra, Università di Ferrara, Ferrara, Italy, \\
    (12) Gran Sasso Science Institute, L'Aquila, Italy
}%

\thanks{This work has been submitted to the IEEE for possible publication.}
}

\maketitle

\begin{abstract}
The BULLKID-DM experiment has been proposed to search for WIMP-like dark matter particles with a mass of 1 GeV or lower and with a WIMP-nucleon interaction cross section down to 10$^{-43}$~cm$^2$. To achieve this goal, the experiment utilizes a detector consisting of a stack of 16 silicon wafers instrumented with phonon-mediated superconducting Kinetic Inductance Detectors (KIDs), to be operated at the underground facility of the Laboratori Nazionali del Gran Sasso (LNGS), Italy.
To reduce the residual background arising from $\gamma$-rays, neutrons, and muons, a cryogenic active anti-coincidence system (veto) operating at mK temperature, surrounding the main detector, is under development. The veto detector consists of segmented scintillating crystals read out by phonon-mediated KID light detectors.
Current status and future developments of the veto system are presented.
In particular, we focus on cryogenic tests of both bismuth germanate (BGO) and gadolinium aluminum gallium garnet (GAGG) crystals, which had been previously chosen as veto candidate crystals according to our preliminary studies on the dependence of their optical properties (i.e. light yield and scintillation decay time) on temperature. We finally present the operation and preliminary tests of a two-module veto prototype with multiplexed readout.
\end{abstract}

\begin{IEEEkeywords}
Article submission, IEEE, IEEEtran, journal, \LaTeX, paper, template, typesetting.
\end{IEEEkeywords}

\section{Introduction}

BULLKID-DM~\cite{colantoni2020bullkid,cruciani2022bullkid,delicato2024data,delicato2024low,delicato2026background,folcarelli2026energy,folcarelli2026projected} aims to search for WIMP-like dark matter particles with masses of 1 GeV or below. The proposed detector consists of a stack of 16 silicon wafers, each instrumented with an array of 145 Kinetic Inductance Detectors (KIDs) to gather phonons produced by particle interactions in a silicon target. Deep grooves machined into the face opposite the KID array divide each silicon wafer into 145 dice, each measuring \(4.5 \times 5 \times 5~\mathrm{mm}^3\). Each KID reads out one die. The main advantage of this segmentation is background rejection through anticoincidence between dice and the definition of a fiducial volume, with the outer dice acting as an active veto. The detector has already demonstrated an energy threshold as low as $200~\mathrm{eV}$, suitable for probing dark matter with masses around $1~\mathrm{GeV}/c^2$, for which the nuclear recoil endpoint in silicon is of order $500~\mathrm{eV}$. The combination of this threshold, a total target mass of approximately $0.8~\mathrm{kg}$, and the large number of multiplexed channels enabling segmentation and background rejection is designed to set new limits on the WIMP–nucleon interaction cross section in the targeted mass range.\\
The experiment will be operated underground at Laboratori Nazionali del Gran Sasso (LNGS) in Italy, where the muon intensity is reduced by six orders of magnitude with respect to the Earth's surface and rock radioactivity has been measured to be lower than in other underground laboratories
~\cite{bucci2009background,belli1989deep,mei2006muon}. . An active anticoincidence detector is being developed to complement the passive shielding and bring the background rate in the target detector down to \(0.1~\mathrm{counts\,kg^{-1}\,day^{-1}\,keV^{-1}}\). This veto consists of a layer of scintillating crystals read out by KID-based light detectors~\cite{lari2026developing}. In this contribution, we provide an overview of this detector and report the latest developments, focusing on millikelvin tests of selected crystals and the operation of a two-module veto prototype with multiplexed readout.
\section{Veto concept}

\begin{figure}
    \centering
    \includegraphics[width=0.9\linewidth]{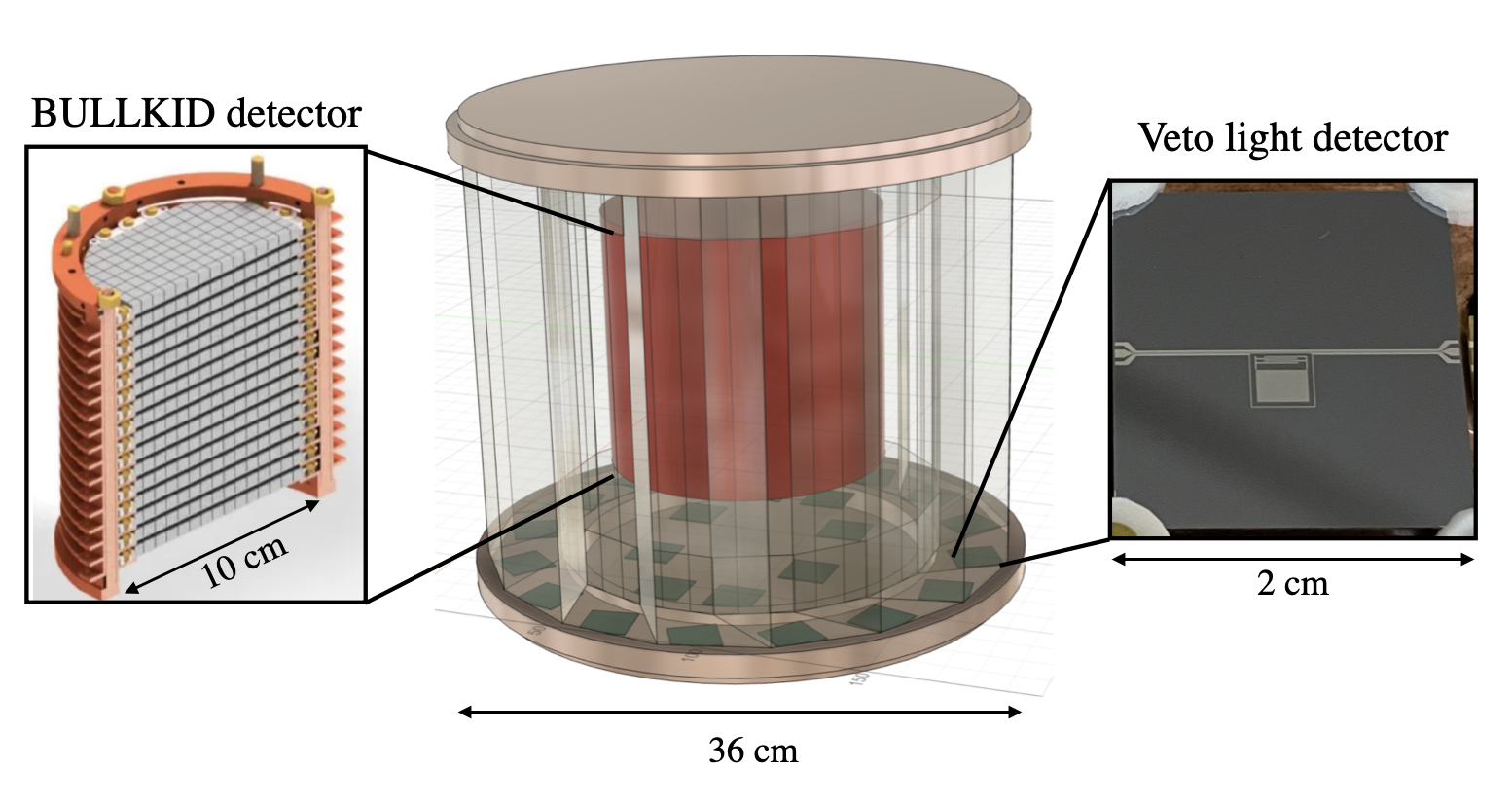}
    \caption{Preliminary layout of the active veto. A segmented layer of scintillating crystals surrounds the BULLKID-DM detector stack, shown in red and in the left inset. The crystals will be read out by approximately 30-60 light detectors, each consisting of a $2\times2~\mathrm{cm}^{2}$ silicon wafer instrumented with KIDs, as shown in the right inset.}
    \label{fig:veolayout}
\end{figure}
The veto must operate at millikelvin temperatures and detect particle
interactions with a threshold below $50~\mathrm{keV}$.

Dense scintillators containing high-atomic-number elements increase the probability of gamma-ray interactions within a compact volume. Crystals containing nuclei with large thermal-neutron capture cross sections can further improve neutron rejection.\\
The proposed layout is shown in Fig.~\ref{fig:veolayout}. Scintillation light produced by particle interactions is read out by an array of KID-based light detectors
consisting of $\sim(2\times2)~\mathrm{cm}^{2}$ silicon wafers instrumented with one or more KIDs, providing a large sensitive area matched to the crystal faces. Photons absorbed in the silicon generate phonons, which are sensed by the KIDs. This light-detector concept has already been demonstrated within the CALDER project~\cite{cardani2021final}.\\
Using KIDs allows us to use the same readout system as the central detector. The current layout requires between 30 and 60 photosensors to ensure efficient light collection. These sensors can be read out using only a few lines through frequency-domain multiplexing.
\section{Crystal selection and background simulations}
Previous studies investigated the light yield and scintillation decay time of candidate crystals down to approximately $20~\mathrm{K}$~\cite{lari2026developing}. These measurements identified bismuth germanate (BGO, $\mathrm{Bi}_{4}\mathrm{Ge}_{3}\mathrm{O}_{12}$) and cerium-doped gadolinium aluminum gallium garnet (GAGG(Ce), $\mathrm{Gd}_{3}\mathrm{Al}_{2}\mathrm{Ga}_{3}\mathrm{O}_{12}:\mathrm{Ce}$) as promising materials. BGO combines high density with efficient $\gamma$-ray attenuation and an increased light yield upon cooling. GAGG exhibited a high light yield at $20~\mathrm{K}$ and contains gadolinium, making it particularly suitable for thermal-neutron capture.\\
Monte Carlo simulations, based on the Geant4 toolkit~\cite{apostolakis2003geant4}, were performed to evaluate the background reduction achievable by replacing part of the passive shielding with these scintillating materials. The reference passive shielding consists of a copper shell with a thickness of $6~\mathrm{cm}$ on the sides and bottom and $12~\mathrm{cm}$ on top. In the simulated active configurations, part of this copper shell is replaced with scintillating crystals.
As previously reported in Ref.~\cite{lari2026developing}, the BGO configuration reduces the $\gamma$-ray background by approximately a factor of 10, while a GAGG veto configuration reduces the neutron background by approximately a factor of 12, relative to the reference passive shielding. Simulations were also performed for muon-induced backgrounds, including secondary particles, showing substantial background suppression for both materials. These results are presented here for the first time.
All these results refer to an energy region of interest of $0.2$-$10~\mathrm{keV}$ in the main detector and a veto energy threshold of $50~\mathrm{keV}$. Table~\ref{tab:vetothresholddependence} reports the background rejection efficiency as a function of the veto energy threshold.
\begin{table}[htbp]
    \centering
    \small
    \setlength{\tabcolsep}{4pt}
    \begin{tabular}{llcccc}
        \hline
        Background & Crystal & $s(50~\mathrm{keV})$
        & \multicolumn{3}{c}{$s(E_{\mathrm{th}})/s(50~\mathrm{keV})$ [\%]} \\
        \cline{4-6}
        & & & $50~\mathrm{keV}$ & $100~\mathrm{keV}$
            & $200~\mathrm{keV}$ \\
        \hline
        $\gamma$-rays & BGO  & 10.07  & 100 & 88 & 66 \\
                     & GAGG & 4.14   & 100 & 85 & 61 \\
        Neutrons     & BGO  & 1.91   & 100 & 89 & 82 \\
                     & GAGG & 12.41  & 100 & 84 & 70 \\
        Muons        & BGO  & 29.05  & 100 & 91 & 81 \\
                     & GAGG & 131.23 & 100 & 81 & 67 \\
        \hline
    \end{tabular}
    \caption{ \parbox{\linewidth}{\normalfont Dependence of the background suppression factor
    $s=N_{\mathrm{passive}}/N_{\mathrm{veto}}$ on the veto energy
    threshold. The third column gives the absolute suppression
    factor at $50~\mathrm{keV}$; the remaining columns give its
    value relative to this reference, expressed as a percentage.
    Event counts refer to the main-detector region of interest.}}
    \label{tab:vetothresholddependence}
\end{table}
Mixed configurations have also been investigated to combine the advantages of the two materials. \\
 The next steps are to test the selected crystals at temperatures of a few tens of millikelvin, representative of the experiment’s operating conditions, and to assess the scalability of the veto architecture. Because the crystals will be installed close to the main detector, their radiopurity is critical. Dedicated measurements are planned at LNGS. In the following sections, we present preliminary measurements of BGO and GAGG coupled to KID light detectors, followed by the operation of a two-module prototype with larger crystals and multiplexed readout.
\section{Millikelvin characterization of BGO and GAGG}
\begin{figure}
    \centering
    \includegraphics[width=0.8
\linewidth]{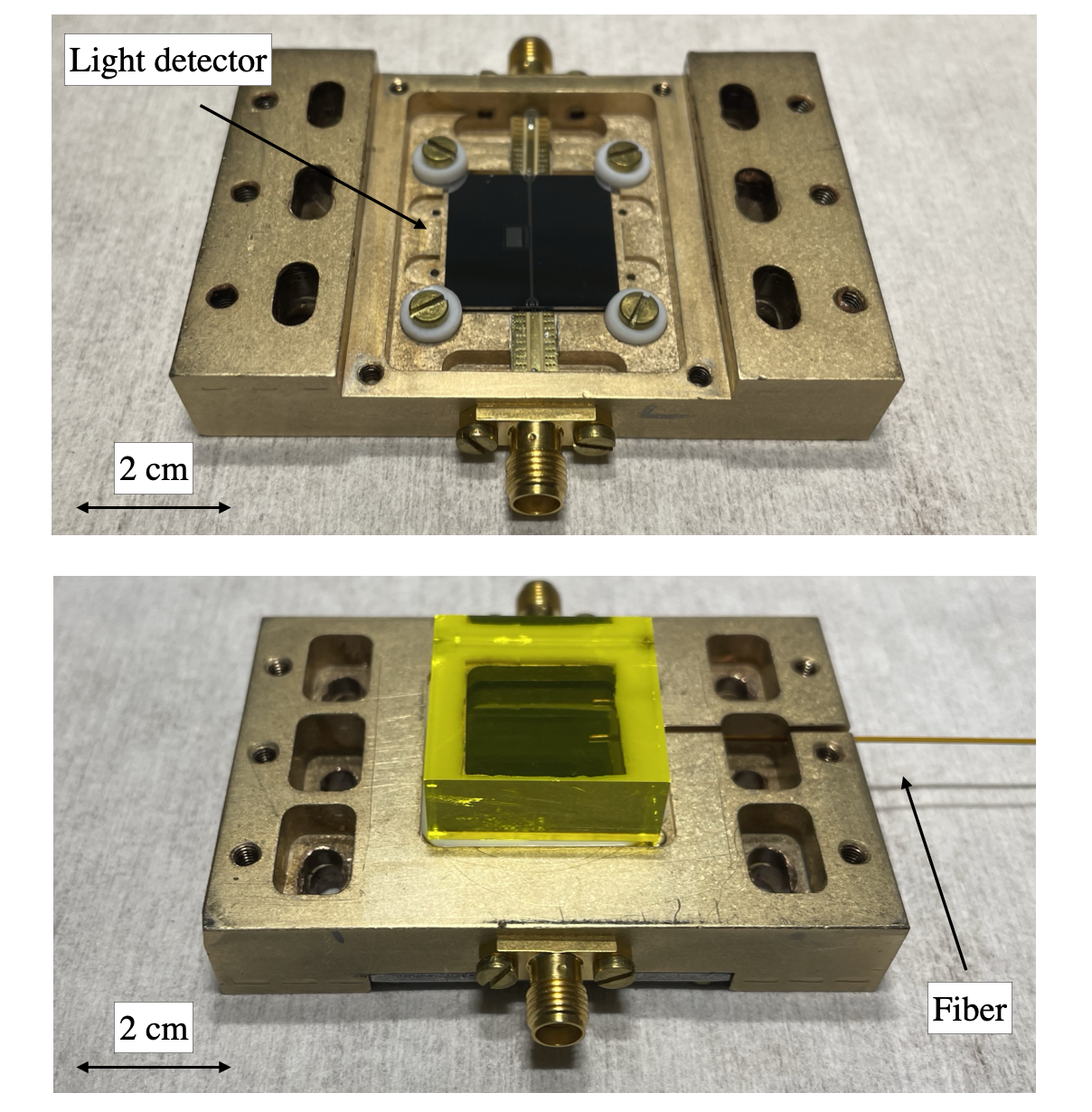}
    \caption{Setup for crystal characterization. Top: the light detector is mounted in the holder and fixed by four PTFE screws. Bottom: the crystal (GAGG in this picture) is placed on the opposite side of the holder, facing the light detector through an aperture. An optical  fiber inserted between the crystal and the light detector delivers LED pulses for calibration.}
    \label{fig:crystal_setup}
\end{figure}

We tested $(2\times2\times1)~{\mathrm{cm}^{3}}$ BGO and GAGG crystals coupled to the KID-based light detector shown in Fig.~\ref{fig:crystal_setup} (top). This device consists of a $20\times20\times0.5~\mathrm{mm}^{3}$ silicon wafer instrumented with a single aluminum KID featuring the same resonator layout as BULLKID-DM detector KIDs. In the light detector the resonator is coupled to a $2~\mathrm{cm}$-long coplanar waveguide (CPW) feedline for readout. We developed two additional designs alongside the standard device used in this test. One features two KIDs on the same wafer (double KID), and the other has a reduced feedline area (narrow-feedline) intended to limit phonon losses in the feedline metal. The characterization and comparison of these designs are ongoing. Resonance frequencies of light detectors span a range $[850-950]~\mathrm{MHz}$.\\
The measurements were performed in an Oxford Instruments Proteox dilution refrigerator at the INFN Pisa cryogenic facility. Dedicated holders were designed to house the crystal and its light detector, as shown in Fig.~\ref{fig:crystal_setup}. Each holder also accommodates an optical fiber used to calibrate the light detector with LED pulses as described in~\cite{del2026lantern}.\\
The assemblies were mounted on the mixing-chamber stage, which reached a temperature of approximately $10~\mathrm{mK}$ during the tests.
For the room-temperature readout, we used the Control and Readout System (CRS), an RFSoC-based electronic board developed by t0.technology~\cite{montgomery2024crs}.
To reconstruct the light-detector signals, we acquired the KID data streams and processed them using a matched filter.
The thresholds quoted in this work are nominal trigger thresholds, set at five times the standard deviation of the filtered noise traces.\\
After calibrating the light detector, we studied the energy response of  KID-crystal system using $662~\mathrm{keV}$ $\gamma$-rays from a $^{137}\mathrm{Cs}$ source placed outside the cryostat.
\begin{figure}
    \centering
    \includegraphics[width=0.9\linewidth]{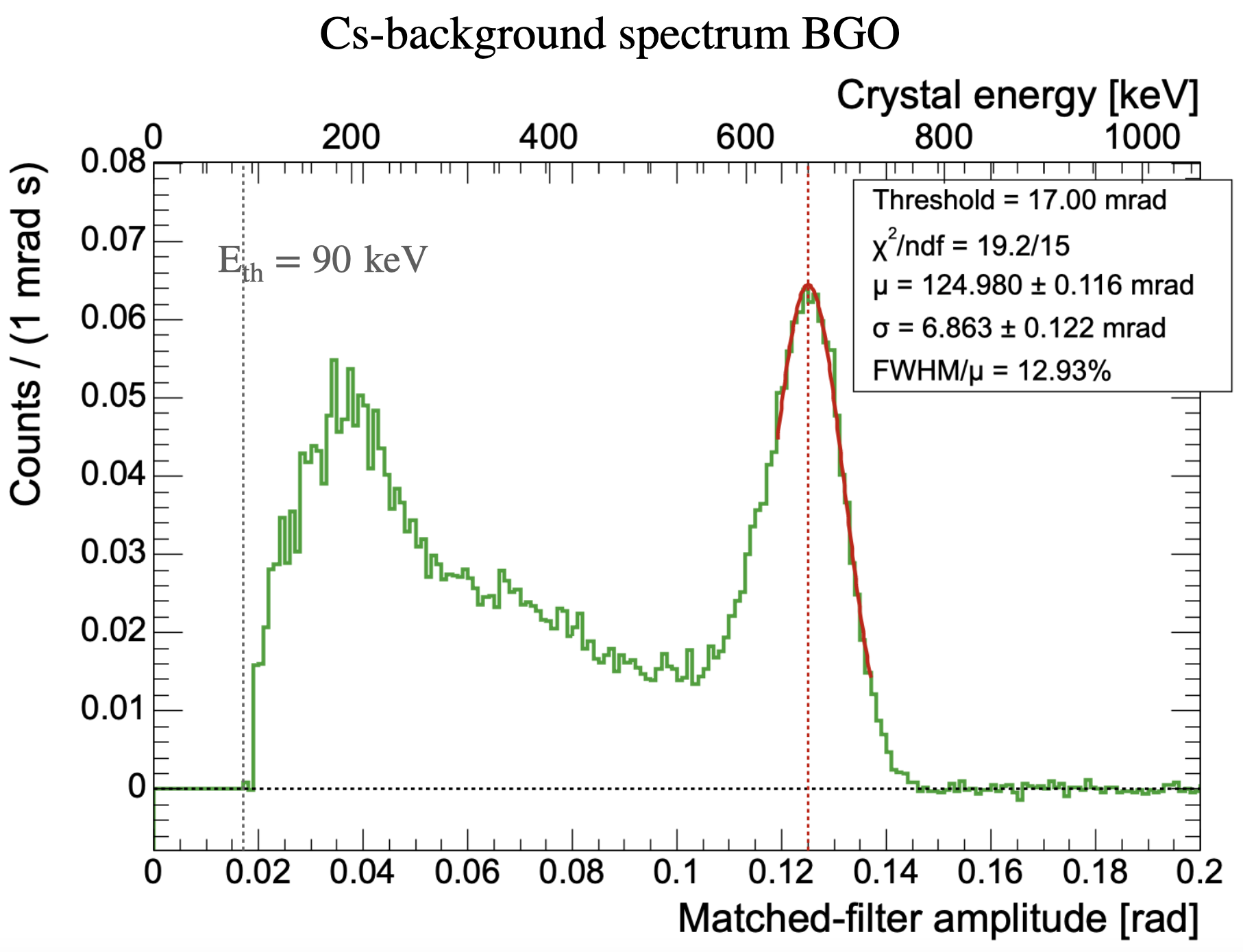}

    \vspace{0.3cm}

    \includegraphics[width=0.9\linewidth]{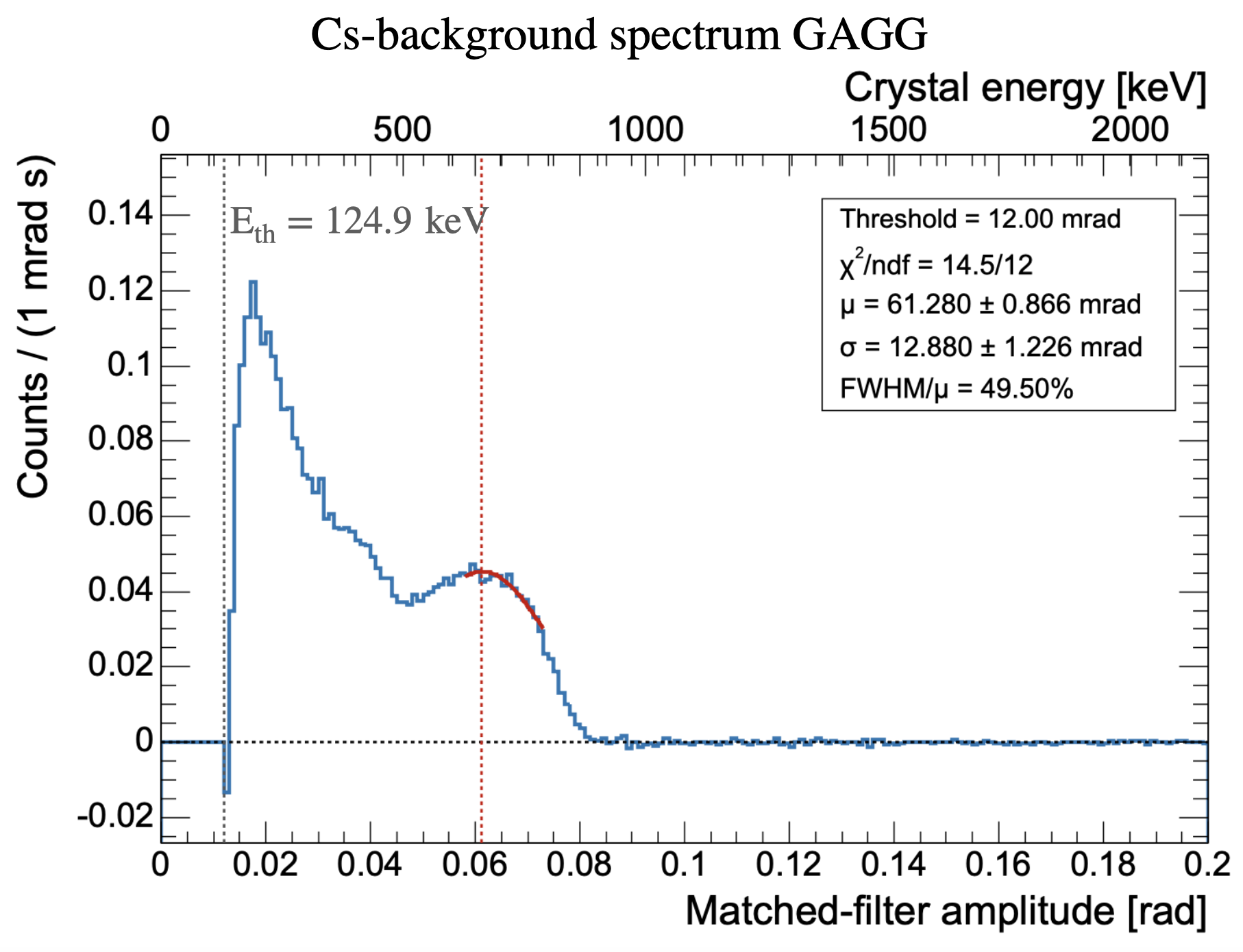}

    \caption{Background-subtracted $^{137}\mathrm{Cs}$ spectra for BGO (top) and GAGG (bottom). The lower horizontal axis shows the reconstructed signal amplitude in radians. The peak position and width were determined with a Gaussian fit, whose results are reported in the inset box. The upper horizontal axis shows the energy deposited in the crystal calibrated using the position of the $662~\mathrm{keV}$ peak. Red dotted line indicates the reconstructed mean of Cs peak while gray dotted line indicates the estimated energy threshold of the  KID-crystal assembly. For each crystal, source and background data were acquired for $10~\mathrm{h}$ each.}
    \label{fig:picchi del cesio}
\end{figure}
Preliminary background-subtracted spectra obtained for BGO and GAGG are shown in Fig.~\ref{fig:picchi del cesio}.\\
BGO exhibits a clear $^{137}\mathrm{Cs}$ photopeak with an energy resolution of $12.9\%$ FWHM and an energy threshold of approximately $90~\mathrm{keV}$. For GAGG, the corresponding values are $49.5\%$ and $130~\mathrm{keV}$, with a detected light yield approximately half that of BGO.
Further measurements with crystals of different geometries will help determine whether the broader photopeak observed in GAGG can be explained by its lower detected light yield alone or whether other effects contribute.
These results point toward a mixed layout with an outer thin GAGG layer (thick enough to efficiently capture thermal neutrons) and an inner bulk BGO layer for gamma-ray energy containment. We are evaluating the neutron shielding performance of this configuration through simulations and planning dedicated experiments to measure its thermal-neutron capture efficiency.\\
 Tests with a similar detector and setup at Sapienza University of Rome yielded a baseline noise of approximately $1.5~\mathrm{mrad}$~\cite{lari2026developing}, compared with $3.5~\mathrm{mrad}$ in the present measurements. At fixed detector responsivity, reaching the same noise level would lower the nominal trigger threshold by a factor of about 2.3. We are improving the setup and investigating the sources of excess noise. We observe spikes in the data streams that we attribute to pulse-tube vibrations, and we are assessing whether these vibrations also contribute to the noise. We are also developing optical simulations which, combined with the light-detector calibration, will allow us to determine the scintillation light yield of both crystals. \\
We also fitted the pulse shapes to extract the scintillation decay time constants of both crystals. The fits yielded two components for each crystal, with the slower component exceeding \(1~\mathrm{ms}\) in both cases.
\begin{table}[htbp]
    \centering
    \small
    \begin{tabular}{lccc}
        \hline
        Crystal & $\tau_1$ [ms] & $\tau_2$ [ms] & $\alpha$ \\
        \hline
        GAGG & $0.0109 \pm 0.0010$ & $2.807 \pm 0.060$
             & $0.7262 \pm 0.0055$ \\
        BGO  & $0.1979 \pm 0.0006$ & $1.2558 \pm 0.0055$
             & $0.7167 \pm 0.0005$ \\
        \hline
    \end{tabular}
    \caption{\parbox{\linewidth}{\normalfont Scintillation decay time constants $\tau_1$ and $\tau_2$
    and fractional contribution $\alpha$ of the fast scintillation component obtained from the pulse-shape fits
    for GAGG and BGO.}}
    \label{tab:scintillation_times}
\end{table}
\section{Two-module prototype with multiplexed readout}
\begin{figure}
    \centering
    \includegraphics[width=0.9\linewidth]{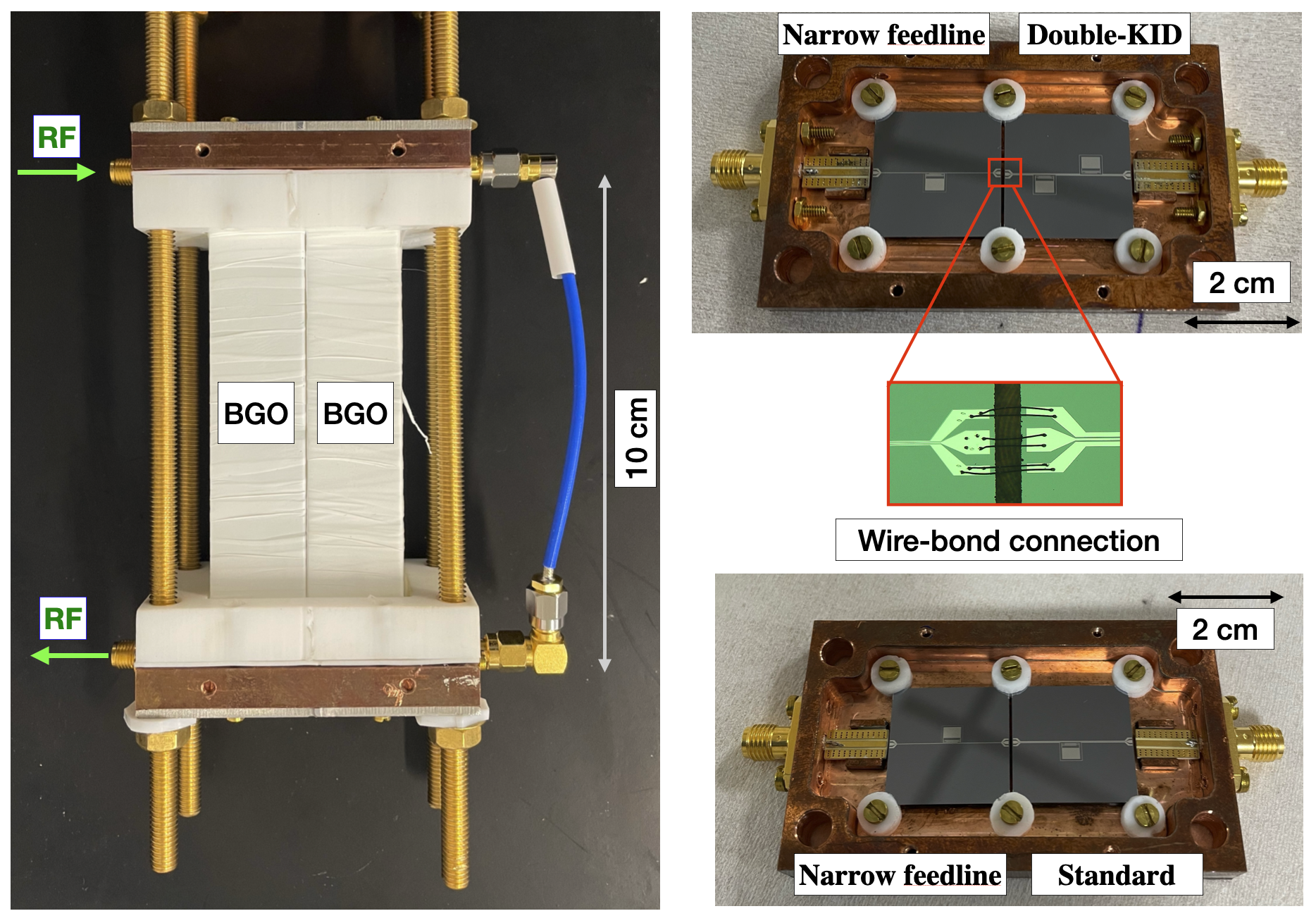}
    \caption{Left: two-module veto prototype consisting of two   BGO crystals read out by four light detectors. To maximize light collection, crystals are wrapped in reflective PTFE tape. Crystals are supported by dedicated copper holders that also house the detectors. The holders and crystals are clamped together in a sandwich assembly using four M6 brass threaded rods with nuts and washers. As shown on the right, each holder contains two light detectors. A continuous readout line connects all four detectors: the feedlines within each holder are connected by wire bonds, while a coaxial cable with SMA connectors links the two holders.}
    \label{fig:veto_prototype}
\end{figure}

To test  larger crystals and demonstrate the scalability of readout, we assembled a prototype, shown in Fig.~\ref{fig:veto_prototype}. It consists of two $(10\times 2.5\times 2)~\mathrm{cm}^3$ BGO crystals read out by four light detectors facing the short end faces of the two crystals.
Dedicated copper holders support the crystals and house two light detectors each. The assembly includes two narrow-feedline devices, one standard device and one double-KID device, for a total of five resonators as shown in Fig.~\ref{fig:veto_prototype}. The feedlines of all four detectors are connected using wire bonds within each holder and a coaxial cable in between. Validating this connection scheme is a key step toward extending the multiplexed readout to the full veto.\\
\begin{figure}
    \centering
    \includegraphics[width=0.9\linewidth]{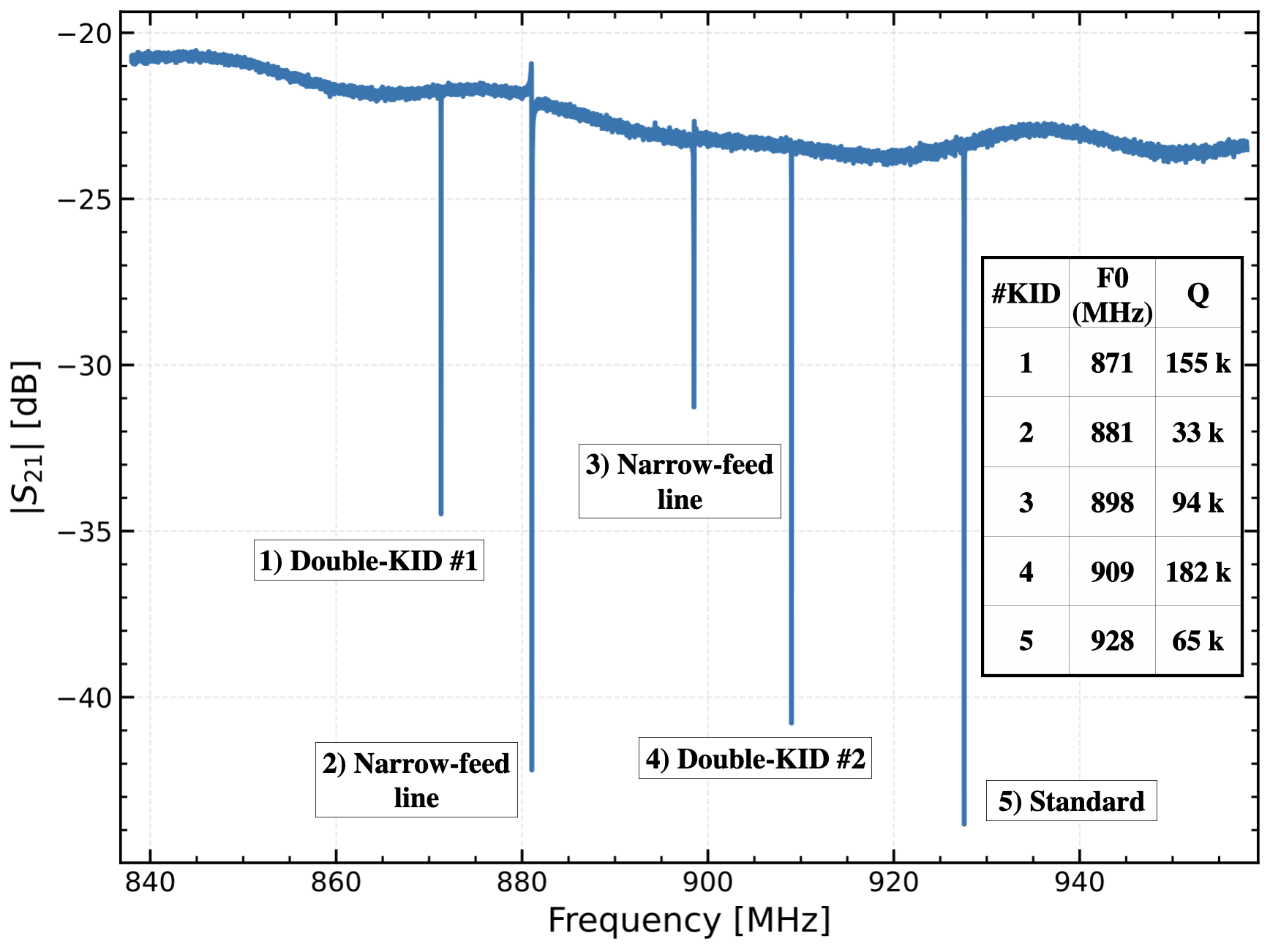}
    \caption{Transmission ($S_{21}$) scan of the two-module prototype, showing all five expected KID resonances. The readout chain includes $80~\mathrm{dB}$ of cryogenic attenuation along the input line, $40~\mathrm{dB}$ of cryogenic amplification and $20~\mathrm{dB}$ of room-temperature amplification, giving a nominal net gain of $-20~\mathrm{dB}$.}
    \label{fig:transmission}
\end{figure}
We first performed a transmission coefficient ($S_{21}$) scan of the complete readout chain. All five expected resonances are clearly resolved in Fig.~\ref{fig:transmission}, confirming that all KIDs can be addressed through the interconnected chips. The resonator with the lowest quality factor has a linewidth below $30~\mathrm{kHz}$. This is sufficiently narrow to accommodate the approximately 60 required resonators within the $750$-$950~\mathrm{MHz}$ readout bandwidth~\cite{ardila2026rfsoc}. The achievable multiplexing factor will also depend on transmission losses along the feedline as more chips are connected. The resonances were assigned to individual KIDs by comparison with their design frequencies and by observing coincident signals in KIDs viewing the same crystal. These results demonstrate the successful operation of the multiplexed readout at the prototype scale.\\
\begin{figure}
    \centering
    \includegraphics[width=0.9\linewidth]{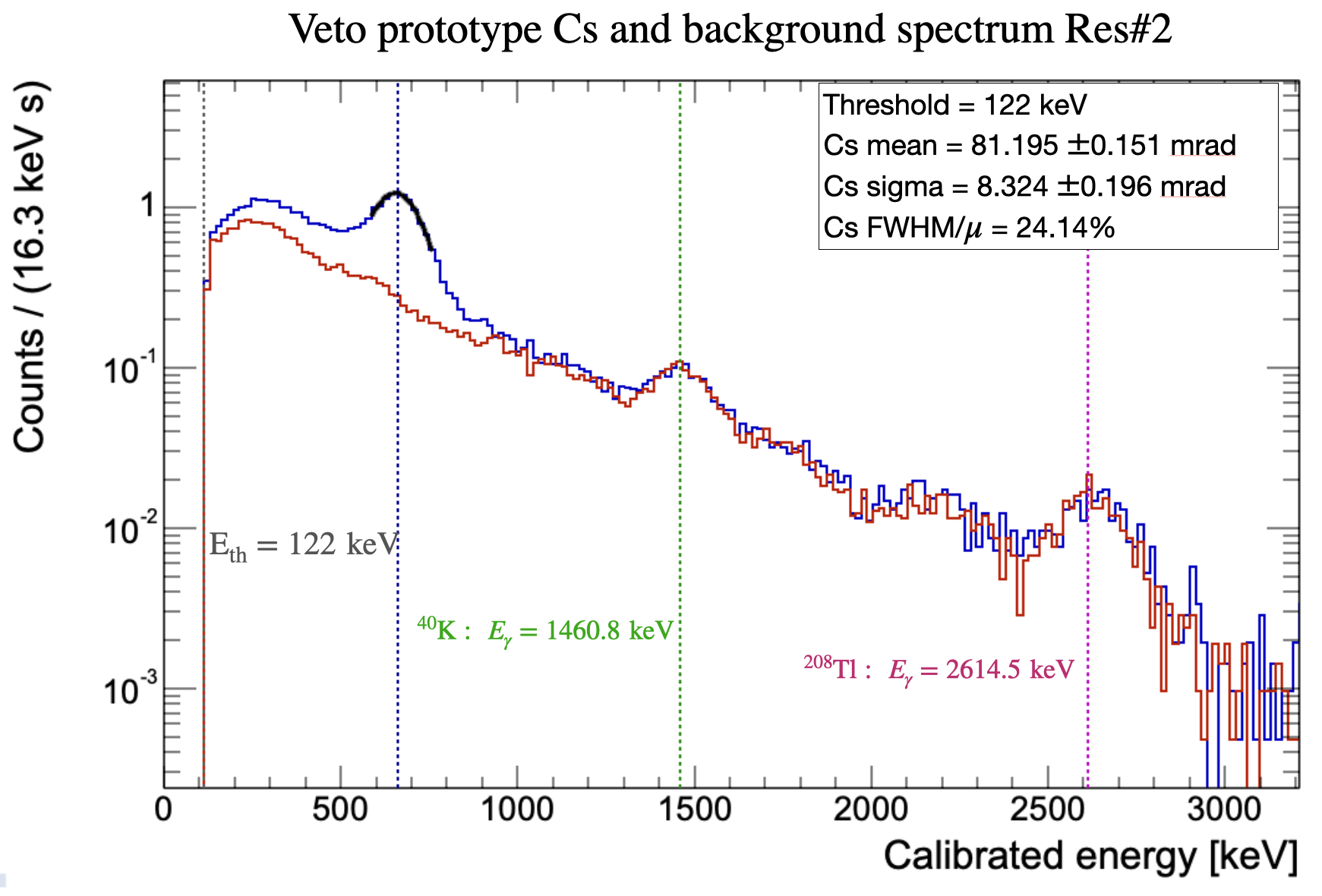}
    \caption{Calibrated energy spectra measured with the best-performing channel of the two-module prototype, with (blue) and without (red) the $^{137}\mathrm{Cs}$ source. The $^{137}\mathrm{Cs}$ photopeak and the environmental $^{40}\mathrm{K}$ and $^{208}\mathrm{Tl}$ $\gamma$-ray peaks are indicated. In  these preliminary tests, the prototype achieved an energy threshold of 122 keV. }
    \label{fig:prototype_spectrum}
\end{figure}
The prototype was then tested with a $^{137}\mathrm{Cs}$ source, and separate environmental background data were acquired for one hour each. In a background acquisition, the rate of these coincident events was approximately ten times that of single-detector events. The origin of the single-detector events will be investigated in further analyses. Figure~\ref{fig:prototype_spectrum} shows the spectra obtained with the best-performing light detector, featuring a reduced-area feedline. The energy scale was calibrated using the $662~\mathrm{keV}$ photopeak. Features consistent with environmental $\gamma$-ray lines from $^{40}\mathrm{K}$ and $^{208}\mathrm{Tl}$ are also visible.\\
Table~\ref{tab:detector_performance_final} summarizes the preliminary results for the individual crystals and the prototype. The photopeak centroids, energy resolutions, and their uncertainties were obtained from fits to the background-subtracted histograms. The source and background histograms were normalized to their respective live times, and the background was subtracted bin by bin with propagation of the Poisson counting uncertainties.
The prototype tests primarily aimed to demonstrate multiplexed readout, and the measured energy resolution and trigger threshold are preliminary. The light-collection geometry differs from that used in the small-crystal tests, and future tests will compare configurations to optimize light collection for the larger crystals.
However, these results are encouraging, with the prototype retaining a nominal trigger threshold  of about $120~\mathrm{keV}$ despite the larger crystal size. Further work will focus on improving the setup and reducing noise to lower the threshold toward the $50~\mathrm{keV}$ target.
\begin{table}

    \centering

    \begin{tabular}{lccc}
        \hline
        Detector & Cs peak & FWHM/$\mu$ & Threshold \\
                 & (mrad)  & (\%)       & (keV)     \\
        \hline
        BGO       & 124.98$\pm$0.12 & 12.93$\pm$0.23 & 90  \\
        GAGG      & 61.28$\pm$0.86 & 49.5$\pm$ 4.7 & 130 \\
        Prototype & 81.19$\pm$0.15 &  24.14$\pm$0.55 & 122 \\
        \hline
    \end{tabular}
    \caption{\parbox{\linewidth}{\normalfont
Preliminary performance of the individual-crystal detectors and the two-module prototype. The prototype values refer to the best-performing channel. Resolutions are quoted at the $^{137}\mathrm{Cs}$ photopeak, and thresholds are nominal trigger thresholds.
}}
\label{tab:detector_performance_final}
\end{table}

\section{Conclusions}
We have presented the BULLKID-DM cryogenic veto project and its latest developments.
We tested BGO and GAGG scintillators with KID light detectors at approximately 10 mK and operated a two-module BGO prototype with multiplexed readout. BGO seems to achieve better preliminary energy resolution as a result of a higher light yield in the small-crystal setup. The larger prototype shows all five expected resonances and reconstructs the $^{137}\mathrm{Cs}$ photopeak with an estimated threshold of 122 keV. These results support the feasibility of a cryogenic scintillating veto with multiple KID light detectors.
\bibliographystyle{IEEEtran}

\bibliography{references}

@article{colantoni2020bullkid,
  title={{BULLKID: Bulky and low-threshold kinetic inductance detectors}},
  author={Colantoni, I and Bellenghi, C and Calvo, M and Camattari, R and Cardani, L and Casali, N and Cruciani, A and Di Domizio, S and Goupy, J and Guidi, V and others},
  journal={Journal of Low Temperature Physics},
  volume={199},
  number={3},
  pages={593--597},
  year={2020},
  publisher={Springer}
}

@article{cruciani2022bullkid,
  title={{BULLKID: Monolithic array of particle absorbers sensed by kinetic inductance detectors}},
  author={Cruciani, A and Bandiera, L and Calvo, M and Casali, N and Colantoni, I and Del Castello, G and del Gallo Roccagiovine, M and Delicato, D and Giammei, M and Guidi, V and others},
  journal={Applied Physics Letters},
  volume={121},
  number={21},
  year={2022},
  publisher={AIP Publishing}
}

@article{delicato2024data,
  title={{The Data Analysis of BULLKID: a monolithic array of particle absorbers sensed by KIDs}},
  author={Delicato, D and Ahmad, A and Bandiera, L and Calvo, M and Cappelli, M and Castello, G Del and Roccagiovine, M Del Gallo and Giammei, M and Guidi, V and Maiello, D and others},
  journal={Journal of Low Temperature Physics},
  volume={215},
  number={3},
  pages={184--192},
  year={2024},
  publisher={Springer}
}

@article{delicato2024low,
  title={{Low-energy spectrum of the BULLKID detector array operated on surface}},
  author={Delicato, D and Ahmad, Abas and Bandiera, L and Calvo, M and Cappelli, M and Del Castello, G and del Gallo Roccagiovine, M and Giammei, M and Guidi, V and Maiello, D and others},
  journal={The European Physical Journal C},
  volume={84},
  number={4},
  pages={353},
  year={2024},
  publisher={Springer}
}

@article{folcarelli2026energy,
  title={{Energy calibration of bulk events in the BULLKID detector}},
  author={Folcarelli, Matteo and Delicato, D and Acevedo-Renter{\'\i}a, A and Ardila-Perez, LE and Bandiera, L and Calvo, M and Cappelli, M and Caravita, R and Carillo, F and Chowdhury, U and others},
  journal={The European Physical Journal C},
  volume={86},
  number={3},
  pages={301},
  year={2026},
  publisher={Springer}
}

@article{delicato2026background,
  title={{Background of the BULLKID detector array operated with moderate shield on surface}},
  author={Delicato, D and Acevedo-Renter{\'\i}a, A and Folcarelli, M and Del Castello, G and Cappelli, M and Ardila-Perez, LE and Bandiera, L and Bonomo, C and Calvo, M and Caravita, R and others},
  journal={arXiv preprint arXiv:2605.02468},
  year={2026}
}

@article{folcarelli2026projected,
  title={{Projected sensitivity to light WIMP-like particles of the BULLKID-DM experiment}},
  author={Folcarelli, Matteo and Acevedo-Renter{\'\i}a, A and Ardila-Perez, LE and Bandiera, L and Calvo, M and Cappelli, M and Caravita, R and Carillo, F and Chowdhury, U and Crovo, D and others},
  journal={arXiv preprint arXiv:2601.13766},
  year={2026}
}

@article{lari2026developing,
  title={{Developing the cryogenic veto for BULLKID-DM experiment}},
  author={Lari, T and del Gallo Roccagiovine, M and Soum-Sidikov, G and Lhuillier, D and Acevedo-Renter{\'\i}a, A and Ardila-Perez, LE and Azzurri, P and Bandiera, L and Calvo, M and Cappelli, M and others},
  journal={IEEE Transactions on Applied Superconductivity},
  year={2026},
  publisher={IEEE}
}

@article{bucci2009background,
  title={{Background study and Monte Carlo simulations for large-mass bolometers}},
  author={Bucci, C and Capelli, S and Carrettoni, M and Clemenza, M and Cremonesi, O and Gironi, L and Gorla, P and Maiano, C and Nucciotti, A and Pattavina, L and others},
  journal={The European Physical Journal A},
  volume={41},
  number={2},
  pages={155--168},
  year={2009},
  publisher={Springer}
}

@article{belli1989deep,
  title={{Deep underground neutron flux measurement with large BF3 counters}},
  author={Belli, P. and others},
  journal={Il Nuovo Cimento A (1965-1970)},
  volume={101},
  number={6},
  pages={959--966},
  year={1989},
  publisher={Springer}
}

@article{mei2006muon,
  title={{Muon-induced background study for underground laboratories}},
  author={Mei and others},
  journal={Physical Review D—Particles, Fields, Gravitation, and Cosmology},
  volume={73},
  number={5},
  pages={053004},
  year={2006},
  publisher={APS}
}

@article{cardani2021final,
  title={{Final results of CALDER: kinetic inductance light detectors to search for rare events}},
  author={Cardani, Laura and Casali, Nicola and Colantoni, Ivan and Cruciani, Angelo and Di Domizio, Sergio and Martinez, Maria and Pettinacci, Valerio and Pettinari, Giorgio and Vignati, Marco},
  journal={The European Physical Journal C},
  volume={81},
  number={7},
  pages={636},
  year={2021},
  publisher={Springer}
}

@article{apostolakis2003geant4,
  title={{GEANT4-a simulation toolkit}},
  author={Apostolakis, J and Arce, P and Burkhardt, H and Chytracek, R and Cosmo, G and Dell'Acqua, A and Folger, G and Giani, S and Ivanchenko, V and Lara, V and others},
  journal={Nuclear Instruments and Methods in Physics Research. Section A: Accelerators, Spectrometers, Detectors and Associated Equipment},
  volume={506},
  number={3},
  pages={250--303},
  year={2003},
  publisher={Elsevier Science Publishing Company, Inc.}
}

@inproceedings{montgomery2024crs,
  title={{The CRS: a scalable full-stack control system for Microwave Kinetic Inductance Detectors}},
  author={Montgomery, Joshua and Avelino, Wellington and Dobbs, Matt and Letang, Joseph and Rouble, Maclean and Savchyn, Sofiia and Smecher, Graeme},
  booktitle={{Millimeter, Submillimeter, and Far-Infrared Detectors and Instrumentation for Astronomy XII}},
  volume={13102},
  pages={309--326},
  year={2024},
  organization={SPIE}
}

@article{del2026lantern,
  title={{LANTERN: Characterization technology for low threshold cryogenic detectors}},
  author={Del Castello, Giorgio},
  journal={IEEE Transactions on Applied Superconductivity},
  year={2026},
  publisher={IEEE}
}

@article{ardila2026rfsoc,
  title={{The RFSoC-based electronics readout system for the BULLKID-DM experiment: sub-GeV dark matter detection with MKIDs}},
  author={Ardila-Perez, LE and Muscheid, T and Gartmann, R and Crovo, D and Cruciani, A and De Lucia, M and Del Castello, G and Delicato, D and Folcarelli, M and Lari, T and others},
  journal={Journal of Instrumentation},
  volume={21},
  number={05},
  pages={C05026},
  year={2026},
  publisher={IOP Publishing}
}

\end{document}